\documentclass[journal]{vgtc}
\usepackage{soul}

\newcommand{\poincare}{Poincar\'e }

\title{An Error-Bounded Lossy Compression Method with Bit-Adaptive Quantization for Particle Data}

\author{
  \authororcid{Congrong Ren}{0009-0006-6285-7271},
  \authororcid{Sheng Di}{0000-0002-7339-5256},
  \authororcid{Longtao Zhang}{0009-0005-3318-8059},
  \authororcid{Kai Zhao}{0000-0001-5328-3962},
  and \authororcid{Hanqi Guo}{0000-0001-7776-1834}
}

\authorfooter{
   \item
  	Congrong Ren and Hanqi Guo are with The Ohio State University.
    E-mail: \{ren.452\,$|$\,guo.2154\}@osu.edu
   \item
    Sheng Di is with Argonne National Laboratory. Email: sdi1@anl.gov
   \item
    Longtao Zhang and Kai Zhao are with Florida State University. Email: \{lz23h\,$|$\,kai.zhao\}@fsu.edu
}

\abstract{This paper presents error-bounded lossy compression tailored for particle datasets from diverse scientific applications in cosmology, fluid dynamics, and fusion energy sciences. As today's high-performance computing capabilities advance, these datasets often reach trillions of points, posing significant visualization, analysis, and storage challenges. While error-bounded lossy compression makes it possible to represent floating-point values with strict pointwise accuracy guarantees, the lack of correlations in particle data's storage ordering often limits the compression ratio. Inspired by quantization-encoding schemes in SZ lossy compressors, we dynamically determine the number of bits to encode particles of the dataset to increase the compression ratio. Specifically, we utilize a k-d tree to partition particles into subregions and generate ``bit boxes'' centered at particles for each subregion to encode their positions. These bit boxes ensure error control while reducing the bit count used for compression. We comprehensively evaluate our method against state-of-the-art compressors on cosmology, fluid dynamics, and fusion plasma datasets.
}

\keywords{Error-bounded lossy compression, particle data, quantization, k-d trees}

\graphicspath{{figs/}{figures/}{pictures/}{images/}{./}}
\usepackage{tabu}                      
\usepackage{booktabs}                  
\usepackage{lipsum}                    
\usepackage{mwe}                       
\usepackage{mathptmx}                  
\usepackage{amsmath}
\usepackage{amssymb}
\usepackage{multirow}
\usepackage{multicol}
\usepackage{algorithm2e}
\RestyleAlgo{ruled}
\usepackage{float}
\usepackage{enumitem}
\usepackage{thmtools,thm-restate}
\usepackage{tablefootnote}
\usepackage{makecell}
\usepackage{wrapfig}
\usepackage{cite}
\usepackage{color}
\usepackage{mathtools}
\DeclarePairedDelimiter\ceil{\lceil}{\rceil}

\definecolor{ferngreen}{rgb}{0.31, 0.47, 0.26}
\definecolor{mynavy}{rgb}{0.259, 0.490, 0.616}
\definecolor{mygray}{rgb}{0.5, 0.5, 0.5}

\begin{document}

\firstsection{Introduction}
\label{sec:intro}

\maketitle

Particle data, which represent discrete entities in scientific simulations such as cosmology~\cite{almgren2013nyx,habib2016hacc,heitmann2019hacc,heitmann2016mira}, fluid dynamics~\cite{scivisContest, sun2019study}, molecular dynamics~\cite{anderson2020hoomd,joshi2021review}, and plasma physics~\cite{verboncoeur2005particle}, exhibit a rapidly increasing volume of data points, with some datasets even reaching trillions~\cite{habib2016hacc}, as computational capabilities expand. The substantial volume of particle data presents considerable challenges in conducting subsequent visualization tasks and analysis, including particle tracking~\cite{chenouard2014objective,meijering2012methods}, statistical analysis~\cite{lista2017statistical,nevcas2012gwyddion}, parameter sensitivity analysis~\cite{marshall2018modeling,singh2006numerical}, and particle-based rendering~\cite{sakamoto2007particle,sims1990particle}. Reducing the storage size of particle data involves meeting several specific criteria. First, ensuring high accuracy in size-reduced data is essential in various applications. For example, in cosmology data, halo detection~\cite{oppenheimer2001direct} demands preserving the relative positions of particles after data reconstruction, typically within a value-range-based relative error bound of $10^{-4}$~\cite{tao2017depth}. Low accuracy of reconstructed data causes failure of tasks, including checkpoint/restart~\cite{cappello2019use}. Second, preserving the positions of all particles is imperative for numerous tasks, such as molecule collision detection~\cite{turk1989interactive} and anomaly detection~\cite{angelov2014anomaly}, making some data reduction methods, such as resampling, unfeasible. Consequently, error-bounded lossy compression has emerged as a direct strategy to mitigate the challenges posed by the vast sizes of scientific data~\cite{ainsworth2020multilevel,lakshminarasimhan2011compressing,liang2022sz3,lindstrom2014fixed}, offering a promising approach for reducing particle storage while retaining all particles.

We are motivated to develop a lossy compression technique explicitly tailored for particle position data, ensuring error control and the preservation of all particles.
We identify two challenges encountered in this endeavor. First, particles are irregularly distributed over space, and the storage ordering of particles does not inherently denote spatial coherence among them, leading to difficulties in predicting particle positions. Most existing studies on lossy compression predominantly concentrate on regular grid data, thus necessitating novel approaches for particle datasets. Second, the trade-off between compression ratio and preserved information is notably more pronounced in particle data compression due to the subsequent tasks involved. Although some efforts have been made to adapt regular grid-based error-bounded compression methods to particle data, these approaches typically yield limited compression ratios, up to 10$\times$ under the relative error bound of 0.001~\cite{jin2020understanding,tao2017depth}. Additionally, specific methodologies that achieve higher compression ratios are primarily oriented toward rendering purposes and result in the loss of particles~\cite{hoang2023progressive}, and thus unsuitable for analyses on or related to the behavior of individual particles or relationships and interactions among them.

This work aims to introduce an error-bounded lossy compression technique for particle position data without reducing the number of particles. Drawing inspiration from the quantization encoder in SZ~\cite{tao2017significantly}, we adaptively select the numbers of bits to encode the quantization codes of particles lossily. Specifically, we initiate the process by partitioning the particles into subregions containing a predetermined number of particles such that the small size of a subregion implies a high density of the particles within the subregion.
Subsequently, we create a box for each subregion,
namely a \textit{bit box}, centered at one of the particles. This box encompasses all particles within the subregion and helps to encode their positions using the smallest number of bits possible, ensuring that a user-specified error bound is satisfied when the center predicts the positions.
Since most particles are within high-density subregions and these subregions have smaller bit boxes, the majority of particles can be encoded by a few bits. We then iteratively remove the bit boxes, along with the particles they contain, one by one, and update the bit boxes intersecting with the removed box until all particles are removed.

We evaluate our method by conducting a comparative analysis with state-of-the-art error-bounded lossy compressors. Specifically, we evaluate SZ2~\cite{liang2018error} and SZ3~\cite{liang2022sz3} with various predictor settings, which are widely used in compressing N-body cosmology data with high accuracy and compression ratios. Additionally, we include MDZ~\cite{zhao2022mdz}, a compression tool designed explicitly for molecular dynamics data, in our evaluation. The baseline comparison results indicate that our approach achieves a higher compression ratio under identical error bounds, as well as superior rate-distortion performance across datasets from diverse application domains, including fluid dynamics, cosmology, and plasma physics. Precisely, our method attains a signal-to-noise ratio (PSNR) of 70.8016 dB with a bit rate of 2.53 bits per particle (BPP) for an instance with 197,761 particles from a time-varying ensemble fluid dynamics dataset~\cite{scivisContest}. In comparison, the maximum PSNR achieved by other compressors under the same BPP setting is 24.7926 dB. Furthermore, given that our method involves reordering particles by grouping them into subregions besides the adaptive lengths of bit boxes, we perform an ablation study to break down the effects of the reordering strategy and the adaptive bit boxes. The results reveal that our reordering strategy enhances the performance of SZ3, while the adaptive bit box lengths also significantly contribute to the overall efficacy of our approach in some cases.

We make the following contributions:
\begin{itemize}
\setlength{\parskip}{0pt}
    \item A framework for compressing particle positions with user-specified pointwise bounded error;
    \item An adaptive and dynamic strategy of bit size selection for quantization codes;
    \item A comprehensive evaluation including baseline compression and ablation study of our method with state-of-the-art compressors.
\end{itemize}

\section{Related Work}
\label{sec:related}

We summarize the relevant literature concerning error-bounded lossy compression, elucidate the disparities between particle data and point cloud compression, and review both domains.

\subsection{Error-bounded Lossy Compression}
Compression techniques can be broadly classified into lossless and lossy compression. While lossless compression is generally not suitable for extensive scientific data due to its limited compression ratios (typically not much better than 2:1 for floating-point~\cite{zhao2021optimizing}), lossy compression in scientific data achieves substantially higher compression ratios while maintaining quality. Lossy compression can be further divided into error-bounded and non-error-bounded categories, depending on whether the pointwise error is constrained within user-specified error bounds. This paper focuses on error-bounded lossy compressors, providing greater control over the distortion of compressed data.

The majority of existing lossy compression techniques are tailored for regular grid data. These techniques encompass prediction-based, transform-based, dimension-reduction-based, and neural-network-based approaches. Prediction-based compressors, exemplified by SZ~\cite{liang2018error,liang2022sz3,tao2017significantly,zhao2021optimizing,zhao2020significantly}, FPZIP~\cite{lindstrom2006fast}, MGARD~\cite{ainsworth2020multilevel}, and ISABELA~\cite{lakshminarasimhan2011compressing}, utilize predictors (such as the Lorenzo predictor) to estimate unknown data points based on known information. Transform-based compression methods (e.g., ZFP~\cite{lindstrom2014fixed} and SPERR~\cite{li2023lossy}) convert original data into sparsely-distributed coefficients, facilitating easier compression. Dimension-reduction-based compression techniques (e.g., TTHRESH~\cite{ballester2019tthresh}) compress data by dimensionality reduction methods, such as higher-order singular value decomposition (HOSVD). In recent years, neural networks have gained widespread adoption for reconstructing scientific data, including autoencoders~\cite{liu2021exploring, zhang2021multi}, superresolution networks~\cite{wurster2022deep, han2021stnet}, and implicit neural representations~\cite{xie2022neural, lu2021compressive, weiss2022fast, martel2021acorn, sitzmann2020implicit}. However, most neural-network-based compressors lack explicit pointwise error control for scientific applications. These compression algorithms have also inspired compressors for irregular data, such as unstructured mesh data~\cite{ren2023prediction}.

It is challenging to apply existing lossy compressors to particle position data due to the irregular structure of such data. The spatial relationships of data points in regular grid data are implicitly encoded by the in-memory indices of the data points, i.e., we know the neighbors of a point by increasing or decreasing its index along one dimension. Differently, the indices of particles do not inherently convey the spatial coherence of particles: two particles with adjacent indices may be distant from each other, and conversely, particles with distant indices may be close in space. However, existing lossy compression algorithms heavily rely on the spatial coherence of data points. For instance, SZ3 predicts the value of a data point based on its neighboring data points; ZFP divides the input data into fixed-size blocks (typically $4\times 4\times 4$ blocks) and compresses every block independently. The lack of implicit spatial structure in input particle data makes these compressors hard to be applied. Thus, we propose a bit box reduction algorithm to establish spatial relationships among particles based on proximity, enabling effective compression, which will be described later.

\begin{table}[!t]
  \caption{Categorization of particle and point cloud compression.}
  \includegraphics[width=\linewidth]{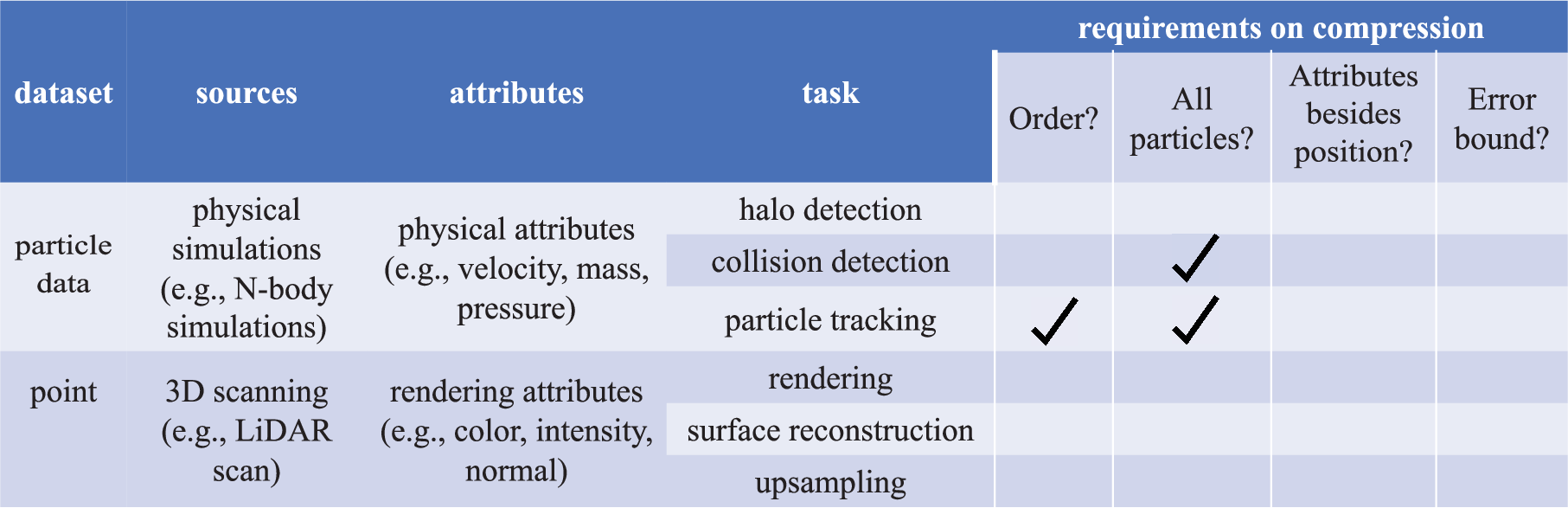}
  \label{tab:related_work}
\end{table}

\subsection{Particle and Point Cloud Compression}
We review existing literature on compression methods for particle and point cloud data, clarifying (1) the similarities and differences between particle and point cloud data and (2) distinctions among compression techniques for positions of particle or point clouds. A brief summary is shown in~\Cref{tab:related_work}.

\textbf{Particle data and point cloud data}. Although particle data and point cloud data are both forms of discrete representations of physical phenomena, they differ in their sources, attributes, and associated tasks. First, particle data typically represents discrete entities in a simulated physical environment. Typical examples include galaxies, star clusters, and molecules generated by N-body simulations~\cite{bagla2005cosmological}. In contrast, point cloud data represents the surfaces of 3D objects and is usually obtained from 3D scanning devices or geometric models, such as a collection of bounce-off positions of laser pulses emitted by a Light Detection and Ranging (LiDAR) sensor~\cite{wiman2009fast} on the surface of an object. Second, particle data is typically characterized by physical attributes such as position, velocity, mass, pressure, and temperature. Besides positions, point cloud data typically contains attributes related to rendering, such as RGB(A) color, intensity, and surface normal. Third, follow-up tasks for point cloud data may involve rendering, surface reconstruction, and super-resolution, where preserving all particles is unnecessary. Conversely, some of followed-up tasks for particle data, such as halo detection in cosmological data and particle collision detection in molecular data~\cite{turk1989interactive}, rely on the attributes of all particles and thus require the preservation of the entire particle set.

\textbf{Overview of particle and point cloud compression}. Following the difference of subsequent analysis tasks of particle and point cloud data, compression techniques can be categorized based on whether a method (1) recovers a complete set or a subset of particles/points in decompression, (2) compresses position only or position with all other attributes on particles/points, (3) preserves the index order of the particles/points, and (4) constrains pointwise error by user-specified bound. The error bound demanded by users could vary by datasets and tasks~\cite{cappello2019use}. Our proposed error-bounded method preserves only the positions of the entire set of particles without preserving the order.

\textbf{Particle compression}.
Most error-bounded compression techniques for particle data applied those designed initially for regular grid data to particle data, which preserves all the particles in their initial order. Additionally, one can choose the attributes to compress. However, these compression techniques have not yet achieved high compression ratios. For instance, Jin et al.~\cite{jin2020understanding} conducted a comprehensive empirical evaluation of GPU-based versions of two error-bounded lossy compressors, SZ and ZFP, using a real-world cosmology particle dataset generated by Hardware/Hybrid Accelerated Cosmology Code (HACC)~\cite{zg3m-8j73-19}.
They achieved the highest compression ratio of 4.25$\times$ under absolute error bounds of 0.005 and 0.025 within a volume of $(0.36\mbox{ Gpc})^3$.
Tao et al.~\cite{tao2017depth} assessed SZ, ZFP, and their improved CPC2000~\cite{omeltchenko2000scalable} on two real-world N-body simulations generated from HACC and Accelerated Molecular Dynamics Family (AMDF) code~\cite{perez2009accelerated}, achieving compression ratio of approximately 10$\times$ under relative error bound of 0.0001. A few studies focusing specifically on particle data compression have emerged, such as MDZ designed for molecular dynamics data by Zhao et al.~\cite{zhao2022mdz}. They specifically dealt with particle data comprising multiple snapshots at multiple timesteps with particular patterns (e.g., zigzag and stairwise patterns) over spatial and characteristics (e.g., high similarity) over temporal dimensions. This approach differs from our work, as we adaptively modulate our approach by particle density without any assumptions about the particle distribution over space or time.

Other studies have specifically designed compressors for particle data but lacked explicit user-specified error control. Devillers and Gandoin~\cite{devillers2000geometric}
proposed a k-d tree-based coder, namely the DG coder (implemented by Google Draco~\cite{draco}), which partitions particles using a k-d tree by splitting a (sub)region in the middle along one of the dimensions until each subregion contains only one particle. The center of its subregion determines the reconstructed position of a particle. Thus, the pointwise position error depends solely on the density around the particle. Hoang et al.~\cite{hoang2023progressive} enhanced the DG k-d-tree-based coder with improved splitting mechanisms (a hybrid of odd-even split and k-d split), traversal strategies (adaptive and block-adaptive traversals), and coding schemes (binary and odd-even split context schemes), and achieved progressive compression and decompression. Although they provided an upper bound of the pointwise error, users have no explicit method to define error bounds. Additionally, their progressive scheme only partially reconstructs the whole set of particles, which is impractical for subsequent tasks such as halo detection~\cite{oppenheimer2001direct}.

\textbf{Point cloud compression}. Traditional and deep-learning-based compression methods for 3D point cloud data have been reviewed by Cao et al.~\cite{cao20193d} and Quach et al.~\cite{quach2022survey}, respectively. Similar to our own method, some approaches also leverage hierarchical data structures like k-d trees to partition the point cloud space into smaller regions with varying levels of detail. For instance, a standardization effort by the Moving Picture Experts Group (MPEG), MPEG Geometry-based Point Cloud Compression (G-PCC)~\cite{schwarz2018emerging}, constructs an octree splitting space into sub-cubes on the midpoint first and represents each point by quantized relative position to the origin of its corresponding sub-cube. This compression method does not explicitly constrain pointwise error, although the reconstruction quality can be inferred from the number of bits used to encode the quantized relative position. Another approach, VoxelContext-Net~\cite{que2021voxelcontext}, applies the same splitting strategy but uses the center as the representative of a cube. It predicts the symbols of non-leaf nodes and introduces a coordinate refinement module to enhance the fidelity of the decompressed point cloud, unlike our method, where the length of quantization codes is dynamically adjusted based on particle density to prioritize compression ratio.

An alternative way to compress point clouds is to use neural representations. This method involves training neural networks using neural field formulations to capture local and global patterns and structures within point clouds. Rather than saving the positions of individual points within a point cloud, the parameters and architecture of the neural network, which require less storage space, are stored instead. See the study by Zeng et al.~\cite{zeng2023self} for a survey. However, this approach is unsuitable for tasks demanding error control at high precision or necessitating the preservation of the particle count.
\section{Preliminary}
\label{sec:preliminary}

\begin{figure}
    \centering
    \includegraphics[width=\linewidth]{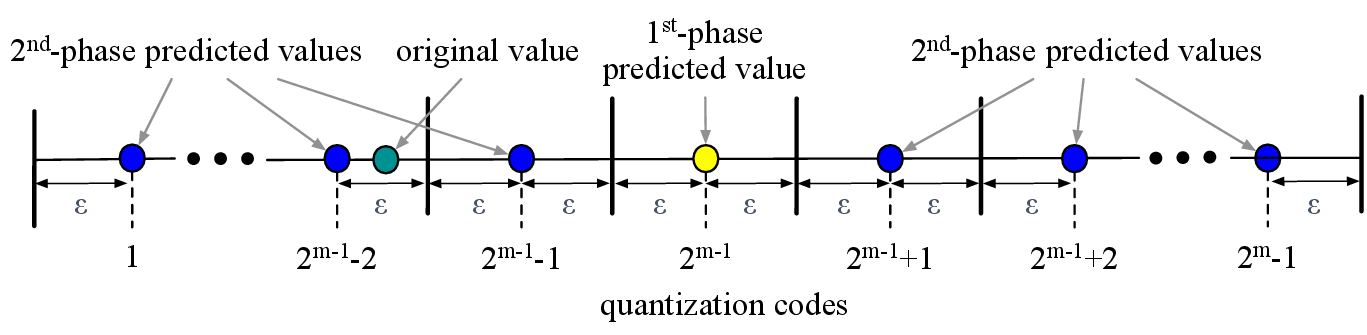}
    \caption{Linear-scaling quantization. The quantization code is $2^{m-1}-2$ for the value in this example. Image reproduced from Figure 2 in~\cite{tao2017significantly}.}
    \label{fig:quantization}
\end{figure}

We investigate prediction-based lossy compression methods with $\epsilon$ as a user-defined \textit{(absolute) error bound}, ensuring $|p_{i,d}-\hat{p}_{i,d}|\leq\epsilon$ for any particle $p_i$ and dimension $d\in\{x,y,z\}$ in the dataset to be compressed, where $p_{i,d}$ and $\hat{p}_{i,d}$ are the original and the decompressed coordinates of $p_i$ along dimension $d$, respectively.
Alternatively, one can also specify $\xi$ as a \textit{relative error bound} to guarantee $\frac{|p_{i,d}-\hat{p}_{i,d}|}{\Delta_{\max}}\leq\xi$,
where $\Delta_{\max}$ represents the maximum range of particle coordinates across all dimensions. We particularly emphasize linear-scale quantization, which serves as the foundation for the quantization method in our algorithm. A prediction-based lossy compressor such as SZ3~\cite{liang2022sz3} usually comprises four primary phases: prediction, quantization, encoding, and lossless compression, which are discussed in detail below.

\textbf{Prediction}. The prediction phase employs interpolation or transformation techniques to predict values of neighboring data points. Commonly used predictors include Lorenzo predictor~\cite{ibarria2003out}, regression predictor, linear interpolation, and cubic spline interpolation~\cite{liang2022sz3}. If the predicted value falls within the error bound, then we say the value is \textit{predictable} and further represent it as a quantization code in the subsequent quantization phase. Otherwise, the value is regarded as \textit{unpredictable} and must be stored in a lossless manner.

\textbf{Quantization}. The quantization phase aims to convert predictable values into quantization codes while controlling errors. Our quantization method draws inspiration from linear-scale quantization, illustrated in~\Cref{fig:quantization} as a 1D example. For a data point whose original value is shown as the green dot, we first obtain its predicted value (depicted as the yellow dot) from the Prediction phase, also referred to as the \textit{$1^{st}$-phase predicted value} (denoted by $pv$).
Suppose the number of bits we use to encode the data point is denoted by $m$, which will be determined by our adaptive strategy in the next section. The linear-scale quantization method then generates $2^m - 1$ successive intervals, each with a length of $2\epsilon$. The $2^{m-1}$-th interval is centered at the 1st-phase predicted value, while the midpoints for other intervals, termed \textit{$2^{nd}$-phase predicted values} (illustrated as blue dots in Figure 2), are determined. A data point whose original value falls within one of these intervals is predictable, and its quantization code is determined by the index of the interval it falls within. The range of a predictable value is
\begin{equation}
    [pv-2\epsilon(2^{m-1}-0.5), pv+2\epsilon(2^{m-1}-0.5)].
    \label{eq:predictable_range}
\end{equation}
During the dequantization phase of the decompression process, the corresponding \textit{$2^{nd}$-phase predicted value} is used to represent the value of the data point, which deviates from the original value by an absolute distance of no more than $\epsilon$.

\textbf{Encoding}. After the initial two phases, the original data is transformed into two sets as bitstreams: quantization codes (for predictable values) and floating-point numbers (for unpredictable values). This phase encodes them by customized variable-length coding techniques to compactly represent the bitstreams. We use Huffman encoding~\cite{huffman1952method} together with R-index based sorting~\cite{omeltchenko2000scalable,tao2017depth}.

\textbf{Lossless Compression}. Finally, the encoded data is further compressed using off-the-shelf lossless compressors such as ZSTD~\cite{zstd}.
\section{Our method}
\label{sec:method}

\begin{figure*}[!th]
    \centering
    \includegraphics[width=\textwidth]{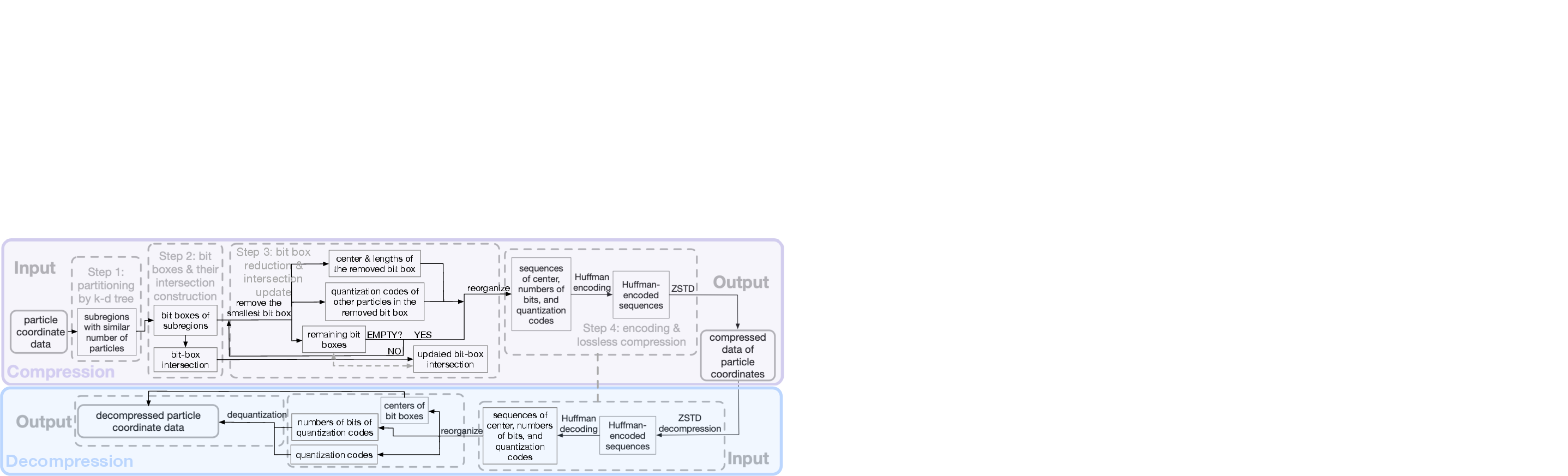}
    \caption{Workflow of our compression and decompression algorithms for particle position data.}
    \label{fig:workflow}
\end{figure*}

The workflow of our compression and decompression algorithms for particle position data is illustrated in~\Cref{fig:workflow}. In the quantization phase, the bounding box encompassing all particles
is partitioned into subregions with similar numbers of particles.
The number of bits to encode particles into quantization codes in each subregion is determined by constructing a bit box for the subregion. A query for bit box intersections is then built. Subsequently, the bit boxes are iteratively eliminated until no particles remain, with simultaneous updates to the bit box intersections. A k-d tree data structure serves as an auxiliary tool in this phase for two purposes: particle partitioning and construction of bit box intersection query. In the encoding phase, R-index based sorting~\cite{omeltchenko2000scalable,tao2017depth} is used to reduce the data layout's entropy and raise the Huffman encoding's compression ratio. Then, Huffman encoding and ZSTD are applied to reduce the data size further.

\begin{algorithm}[!ht]
\scriptsize
\caption{Bit Box Reduction Compression}\label{alg:bit_box_reduction}
\KwData{points: coordinates of a set of particles;\\
\hspace{0.85cm}$\xi$: user-defined relative error bound;\\
\hspace{0.85cm}r: maximum number of points allowed in a subregion}
root $\gets$ Node(\textit{pts} = points, \textit{bb\_of\_bit\_boxes} = \texttt{None},
 \textit{split\_pt} = \texttt{None}, \textit{parent} = \texttt{None}, \textit{left\_child} = \texttt{None}, \textit{right\_child} = \texttt{None})\;
tree $\gets$ KDTree(\textit{root}=root)\;
subregions $\gets$\ tree.\textit{get\_regions\_with\_num\_pts}(r)\;
\For{node $\in$ subregions}{
node.\textit{bb\_of\_bit\_boxes} $\gets$ get\_bit\_boxes(node.\textit{pts}, $\xi$)\;
}
tree.\textit{bottom\_up\_initialize\_bb\_of\_bit\_boxes}()\;
quant\_codes $\gets$ []\;
remaining\_pts $\gets$ points\;
\While{remaining\_pts$\neq\emptyset$}{
curr\_subregion $\gets$ argmin(subregions.\textit{bb\_of\_bit\_boxes})\;
intersected\_subregions $\gets$ tree.\textit{top\_down\_search\_intersection}(curr\_subregion)\;
all\_common\_pts $\gets \emptyset$\;
\For{node $\in$ intersected\_subregions}{
common\_pts $\gets$ get\_common\_pts(curr\_subregions, node)\;
all\_common\_pts.\textit{add}(common\_pts)\;
node.\textit{pts}.\textit{remove}(common\_pts)\;
node.\textit{bb\_of\_bit\_boxes} $\gets$ get\_bit\_boxes(node.\textit{pts}, $\xi$)\;
}
quant\_codes.\textit{add}(curr\_subregion.\textit{bb\_of\_bit\_boxes.center})\;
quant\_codes.\textit{add}(curr\_subregion.\textit{bb\_of\_bit\_boxes.size})\;
quant\_codes += quantize(curr\_subregion.\textit{pts} + all\_common\_pts)\;
quant\_codes.\textit{add}(end\_mark)\;
remaining\_pts.\textit{remove}(curr\_subregion.\textit{pts} + all\_common\_pts)\;
subregions.\textit{remove}(curr\_subregion)\;
curr\_subregion.\textit{bb\_of\_bit\_boxes}=\texttt{None}\;
tree.\textit{bottom\_up\_update\_bb\_of\_bit\_boxes}()\;
}
\textbf{return} quant\_codes
\end{algorithm}

\subsection{Partitioning and quantization}
\label{sec:bit_box}
Without loss of generality, we use 3D particle data as an example to describe our algorithm, which can be easily adapted to 2D particles.
Our quantization phase comprises three steps, as outlined below.

\begin{figure*}[!th]
    \centering
    \includegraphics[width=\textwidth]{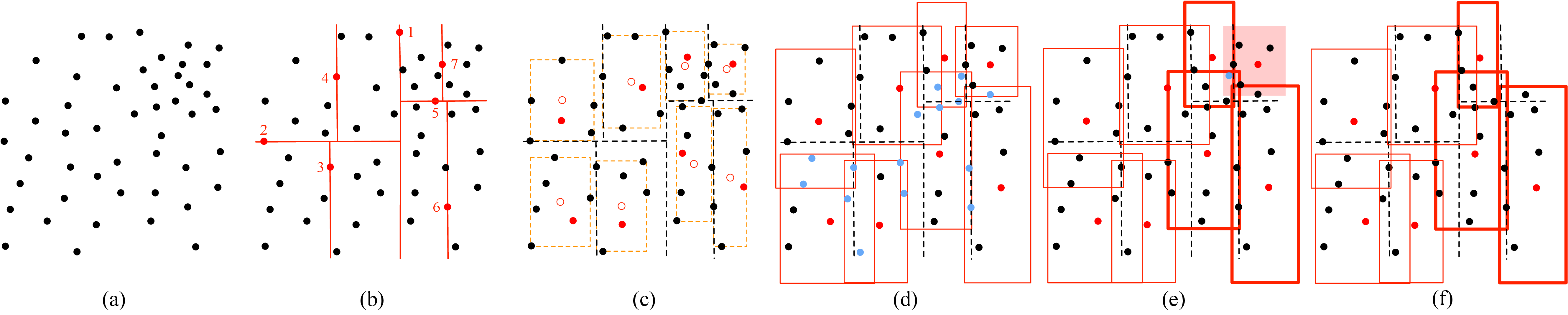}
    \caption{Illustration of our bit box reduction algorithm for particle position compression by a 2D example. (a) A set of particles. (b) Particles are divided into groups via k-d tree splitting at the median. The red lines represent the splitting (hyper)plane, while the red dots indicate the medians. Numbers in red denote the order of splitting.
    The splitting process terminates when each subregion contains no more than $r$ particles, with $r$ being an integer ranging from 6 to 11 in this example. (c) We start with selecting centers for bit boxes. Initially, AABBs, delineated by red dashed lines, are created for the subregions. A particle, indicated by a solid red dot, closest to the center (represented by a hollow red dot) of the AABB, is chosen as the center of the corresponding bit box for the subregion.
    (d) We construct bit boxes for subregions centered at the solid red dots, as indicated in (c), with lengths calculated based on Equations~\ref{eq:num_bits} and~\ref{eq:box_length}. Particles that fall within the overlap of bit boxes are highlighted in blue. (e) The bit box (shown as the solid red box) with the smallest sum of lengths across all dimensions is selected for elimination. The center of the bit box is losslessly stored. In contrast, all other particles inside the bit box, including those initially belonging to other subregions (e.g., the blue dot), are quantized using the numbers of bits determined by~\Cref{eq:num_bits}. All intersecting bit boxes are highlighted by thick lines. (f) All intersecting bit boxes are updated, as removing particles from their respective subregions may reduce the sizes (and potentially change the centers) for these bit boxes.}
    \label{fig:method}
\end{figure*}

\textbf{Initial particle partitioning with k-d traversal}.
We first partition all particles into groups by a k-d tree, where an axis-aligned (hyper)plane passing the median of particles along one single dimension is chosen as the splitting (hyper)plane to separate the current (sub)region into two subregions, as shown in~\Cref{fig:method}(b).
The median on the splitting (hyper)plane is considered to be on the side
with larger coordinates on the dimension. Splitting by median ensures a balanced number of particles between the two resulting children and potentially yields smaller bounding boxes, thereby fewer bits for encoding particles, which will be reasoned later. Notably, the splitting of a subregion terminates when the subregion accommodates no more than a certain number of particles (see~\Cref{fig:method}(b)). We denote the maximum number of particles allowed within a subregion by $r$ and define the ratio of this number to the total number of particles $N$ as $\tilde{r}$ (i.e., $\tilde{r}=r/N$), whose tuning will be discussed in Section \ref{sec:parameter_analysis}.

\textbf{Bit box construction}. We construct a bit box for every subregion with no more than $r$ particles. We first determine the axis-aligned bounding box (AABB), the smallest box encompassing all particles while aligning with the coordinate axes for every subregion (boxes in red dashed lines in~\Cref{fig:method}(c)). The center of a bit box, denoted by $c$, is determined by the particle closest to the center of the AABB.
Note that the center of a bit box (red solid dots in~\Cref{fig:method}(c)) is a particle within the subregion, while the center of an AABB (red hollow circles in~\Cref{fig:method}(c)) of the subregion is not. If we use $c_d$, the coordinate of $c$ along one dimension $d\in\{x, y, z\}$, to predict coordinates along dimension $d$ of all particles in the subregion, then to ensure that all the particles within the subregion are predictable, we need the half-length of the predictable range (i.e., $2\epsilon(2^{m-1}-0.5)$ given by~\Cref{eq:predictable_range}) to be at least $max(c_d - pos\_min_d, pos\_max_d - c_d)$, where $pos\_min_d$ and $pos\_max_d$ are the smallest and largest particle coordinates in the subregion along dimension $d$. This condition determines the smallest number of bits $m_d$ required to encode the coordinates in dimension $d$:
\begin{equation}
    m_d = \ceil{log_2(\frac{max(c_d - pos\_min_d, pos\_max_d - c_d)}{2\epsilon} + 0.5) + 1},
    \label{eq:num_bits}
\end{equation}
where $\ceil{\cdot}$ denotes the ceiling function.

We use different $m_d$ values for distinct dimensions, rather than the same number of bits for all dimensions, to reduce the number of bits needed to encode a particle, particularly for subregions with varied ranges across different dimensions. For example, with $\epsilon=0.005$ set, consider a subregion with an AABB defined by two extreme points $(0,0,0)$ and $(30,0.02,2)$, which has highly nonuniform ranges across three dimensions, with its center located at $(15,0.01,1)$. Assuming the closest particle to the AABB center has coordinates $(14,0.009,1.1)$, then we have $m_x=12$, $m_y=2$, and $m_z=8$ by~\Cref{eq:num_bits}, which requires only 22 bits to represent quantization code of a particle within the subregion. However, if we used the same number of bits for encoding coordinates in all dimensions, we would need 12 bits for each dimension and 36 bits overall to encode a particle, resulting in a substantial overhead for more significant numbers of particles.

We use a bit box to represent the predictable range under $m_d$ of a subregion by setting the length $l_d$ of the bit box in dimension $d$ to be:
\begin{equation}
    l_d = 4\epsilon(2^{m_d-1}-0.5),
    \label{eq:box_length}
\end{equation} 
which is derived by plugging~\Cref{eq:num_bits} into~\Cref{eq:predictable_range}.
Note that the bit box of a subregion is larger than the AABB of the particles in the subregion because the ceiling function in~\Cref{eq:num_bits} raises a value. Thus, a potential intersection and common particles (e.g., blue particles in~\Cref{fig:method}(d)) may exist between two bit boxes.

The rationale behind maintaining a balanced number of particles in two half spaces during splitting is to accommodate the particle density such that we can approach the possibility of encoding particles with minimal bit usage. There is a different strategy to partition particles, that is to split a (sub)region by midpoint, which has been used by several prior studies~\cite{devillers2000geometric,draco,hoang2023progressive,que2021voxelcontext}. Different from this approach which prioritizes spatial uniformity, we achieve an even distribution of particle counts instead to enable adaptive sizing of bit boxes to match the particle density such that particles located in high-density areas are encapsulated by smaller bit boxes and thus encoded by fewer bits. Given that the majority of particles tend to be distributed in high-density areas, our approach effectively reduces the overall number of bits used to encode quantization codes. 

We devise a bit box intersection query that takes a bit box as input and returns all bit boxes intersecting with it by traversing the k-d tree constructed previously for particle partitioning to enhance algorithm efficiency. 
When constructing the k-d tree, each node encapsulates the particles within the represented (sub)region. For instance, the root of the k-d tree comprises all particles, while the two children of the root encompass particles within the two subregions generated by the first splitting. Leaf nodes represent subregions with no more than $r$ particles. After bit box construction, each leaf node in the k-d tree stores corresponding bit box information, including the center and length of its bit box. Instead of bit boxes, the non-leaf nodes of the k-d tree store AABBs. Specifically, a parent of two leaf nodes stores the AABB encompassing the two bit boxes associated with the respective leaf nodes, while any other non-leaf node stores the AABB of AABBs of its two children. The initialization of AABBs for non-leaf nodes is facilitated by a bottom-up strategy. To query the intersecting bit boxes of a given bit box, we conduct box-box intersection tests starting with the root of the k-d tree.
Subsequently, we use a top-down search strategy to locate the leaf nodes representing intersecting bit boxes. Both bottom-up initialization or update and top-down search incur a time complexity of $O(log|B|)$, which is a significantly more efficient strategy compared to exhaustively visiting all bit boxes ($O(|B|)$), where $|B|$ denotes the number of bit boxes.

\textbf{Bit box reduction and bit box intersection update}. In each iteration, we remove the bit box with the smallest $\sum_d m_d$, i.e., the total number of bits to encode one particle in the bit box, along with the sequence of particles within it. An example of such a bit box is highlighted in shadow in Figure \ref{fig:method}(e). The center of the bit box to be removed is losslessly encoded, while all other particles within the bit box, including the common particles (e.g., the blue dot in~\Cref{fig:method}(e)) that are initially assigned to other subregions, are quantized by $m_d$ derived from~\Cref{eq:num_bits}. Since particles within the intersection of two bit boxes are also removed, the remaining bit boxes intersecting with the one to be removed may include fewer particles or even become empty. Consequently, there is potential for these intersecting bit boxes to decrease in size and be encoded with fewer bits $m_d$. Hence, we update both the centers and lengths for the intersecting bit boxes by the new AABBs of remaining particles within the corresponding subregions, as well as the bit box intersection by bottom-up traversal in the k-d tree. For example, the intersecting bit box to the left of the shadowed bit box in~\Cref{fig:method}(e) becomes narrower in~\Cref{fig:method}(f).
Empty bit boxes are hidden in subsequent processes. Let $M$ denote the precision of the original data, where $M=32$ for single and 64 for double precision floating points. If $m_d>M$ for some $m_d$, then instead of storing the quantization codes, we losslessly store the original coordinates of particles in dimension $d$ within the bit box. The pseudocode of our bit box reduction algorithm is presented in~\Cref{alg:bit_box_reduction}.

\begin{figure}[!t]
    \centering
    \includegraphics[width=\linewidth]{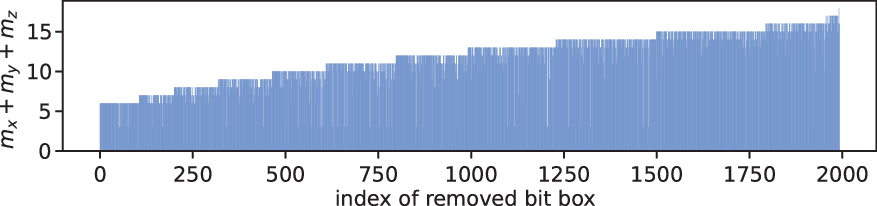}
    \caption{The total number of bits across all dimensions ($\sum_d m_d$) for quantization codes of bit boxes for NYX dataset with $\xi=0.005$ and $\tilde{r}=0.05\%$. $\sum_d m_d$ is not monotonically increasing, and the number of removed bit boxes (i.e., 1,993) is smaller than the total number of bit boxes (i.e., 2,000). See explanation in the last paragraph in~\Cref{sec:bit_box}.}
    \label{fig:num_bits}
\end{figure}

The heuristic used to remove the bit box with the smallest $\sum_d m_d$ is greedy. We chose this heuristic because the density within the chosen bit box is highest given that every subregion contains similar numbers of particles, and thus, more common particles in other subregions can be quantized by fewer bits. Notably, removing all bit boxes to eliminate all particles may not be necessary, as previously removed bit boxes might take all particles within some bit boxes. \Cref{fig:num_bits} plots the $\sum_d m_d$ of removed boxes on the NYX dataset with $\xi=0.005$ and $\tilde{r}=0.05\%$. It is evident that: (1) although we always select the bit box with the smallest $\sum_d m_d$, the $\sum_d m_d$ of the removed bit box does not monotonically increase due to shrinking sizes of intersecting bit boxes; (2) The number of removed bit boxes is slightly smaller than the total number of bit boxes initially constructed (to be exact, 1,993 out of 2,000), as some intersecting bit boxes become empty.

\subsection{Quantized data layout and compression}

\begin{wrapfigure}{LH}{0.21\linewidth}
\centering
\vspace{-\intextsep}
\tiny
\fbox{
\parbox{\linewidth}{
\textbf{seq$_\mathtt{0}$}=$[$\textcolor{violet}{coors of center0},\\
\textcolor{ferngreen}{num\_bits of seq0},\\
\textcolor{mynavy}{quant\_code of point00},\\
\textcolor{mynavy}{quant\_code of point01},\\
...,\\
\textcolor{mygray}{sequence end mark}$]$,\\
\textbf{seq$_\mathtt{1}$}=$[$\textcolor{violet}{coors of center1},\\
\textcolor{ferngreen}{num\_bits of seq1},\\
\textcolor{mynavy}{quant\_code of point10},\\
\textcolor{mynavy}{quant\_code of point11},\\
...,\\
\textcolor{mygray}{sequence end mark}$]$,\\
...,\\
\textbf{seq}$_\mathtt{s}$=$[$\textcolor{violet}{coors of center\_s},\\
\textcolor{ferngreen}{num\_bits of seq$_\mathtt{s}$},\\
\textcolor{mynavy}{quant\_code of point\_\{s,0\}},\\
\textcolor{mynavy}{quant\_code of point\_\{s,1\}},\\
...,\\
\textcolor{mygray}{sequence end mark}$]$
}}
\caption{Data layout of sequences}
\vspace{-\intextsep}
\label{fig:sequences}
\end{wrapfigure}

With the bit box reduction process, we organize information of removed bit boxes into sequences, as depicted in~\Cref{fig:sequences}. Each sequence corresponds to a removed bit box and starts with \textcolor{violet}{the coordinates of bit box center}, stored as lossless floating points. Next, the \textcolor{ferngreen}{numbers of bits} $\{m_x,m_y,m_z\}$ required to encode other particles within the bit box are specified. Given that $m_d$ is constrained by $M$, where $M=32$ or $M=64$, each $m_d$ can be represented by 5 or 6 bits, respectively. The primary segment of a sequence is \textcolor{mynavy}{quantization codes} for particles stored as integers. An \textcolor{mygray}{end mark}, stored as a 32-bit integer, is appended to signify the sequence's conclusion. Once all particles are removed, the data layout of sequences contains all the information necessary to reconstruct the original dataset.

To reduce the entropy of values in the data layout and improve the compression ratio achieved by Huffman encoding, we apply two reordering strategies: sequence reordering and quantization code reordering within each sequence. Sequence reordering is performed based on the total number of bits of quantization codes, i.e., $\sum_d m_d$. Quantization codes within a sequence are then reordered based on their R-indices, following the interleaved pattern proposed by Omeltchenko et al.~\cite{omeltchenko2000scalable}. In this pattern, quantization codes of three coordinates are sorted based on their respective entropy levels. For instance, consider a set of sequences where $m_x$ ranges from 2 to 4 uniformly, $m_y$ ranges from 2 to 9 uniformly, and $m_z$ ranges from 2 to 9 with a single-peak distribution. Among these dimensions, $y$ and $x$ exhibit the highest and lowest entropy, respectively. Consequently, we order quantization codes within the sequences as $(x, z, y)$ to generate R-indices. As an illustration, a particle encoded by quantization codes $(01, 100111, 01010)$ yields an R-index of $0101010100111$. Then, the quantization codes of particles in each sequence are sorted by the R-indices. These reordering procedures aim to increase the frequency of frequently-repeated bit sequences, resulting in smaller bit representations after Huffman encoding. By implementing these two reordering mechanisms, we observed an improvement in compression ratio ranging from 1.05 to 1.1 times in tested datasets (shown later in~\Cref{sec:eval}). Finally, the sequences undergo the Huffman encoding and lossless compression process akin to other prediction-based lossy compressors, as reviewed in~\Cref{sec:preliminary}.

\begin{figure}[!t]
    \centering
    \includegraphics[width=0.8\linewidth]{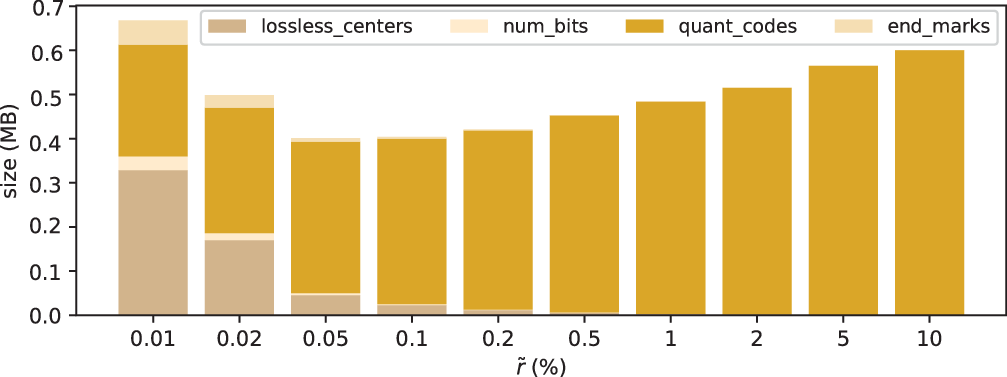}
    \caption{Breakdown of storage cost of data layout on NYX dataset with relative error bound $\xi=0.005$. As the parameter $\tilde{r}$ increases, the sizes for losslessly stored centers, numbers of bits for quantization codes, and end marks also increase, while the total size of quantization codes decreases. The overall size of the layout reaches its minimum at $\tilde{r}=0.05\%$.}
    \label{fig:size_breakdown}
\end{figure}

We examine the storage cost of the data layout. Let $n_{seq}$ denote the number of sequences, $num\_particles_s$ denote the number of particles in sequence $s$, and $m_{s,d}$ denote the number of bits to encode quantization codes of dimension $d$ in sequence $s$. Note that $n_{seq}$ is not equivalent to the number of boxes $|B|$ because particles within some bit boxes might be all removed when eliminating their intersecting bit boxes. The storage cost consists of four components: (1) $3Mn_{seq}$ bits for losslessly stored centers; (2) $15n_{seq}$ bits for $M=32$ and $18n_{seq}$ bits for $M=64$ to record numbers of bits; (3) $\sum_{s=0}^{n_{seq}-1}(m_{s,x}+m_{s,y}+m_{s,z})(num\_particles_s-1)$ bits to store all quantization codes; (4) $32n_{seq}$ bits for end marks. With the maximum number $r$ of particles allowed in each subregion decreases, the number of bit boxes together with $n_{seq}$ increases. At the same time, $num\_particles_s$ and $m_{s,d}$ tend to decrease, which reduces the storage cost for (3) and raises the costs for (1), (2), and (4), as shown in~\Cref{fig:size_breakdown}.
The overall storage cost would reach a minimum at $\tilde{r}=0.05\%$. However, this $\tilde{r}$ does not necessarily indicate the optimal compression ratio because the subsequent Huffman encoding and ZSTD compression reduce the size of different data layouts to varying extents.

\begin{figure*}[!th]
    \centering
    \includegraphics[width=\textwidth]{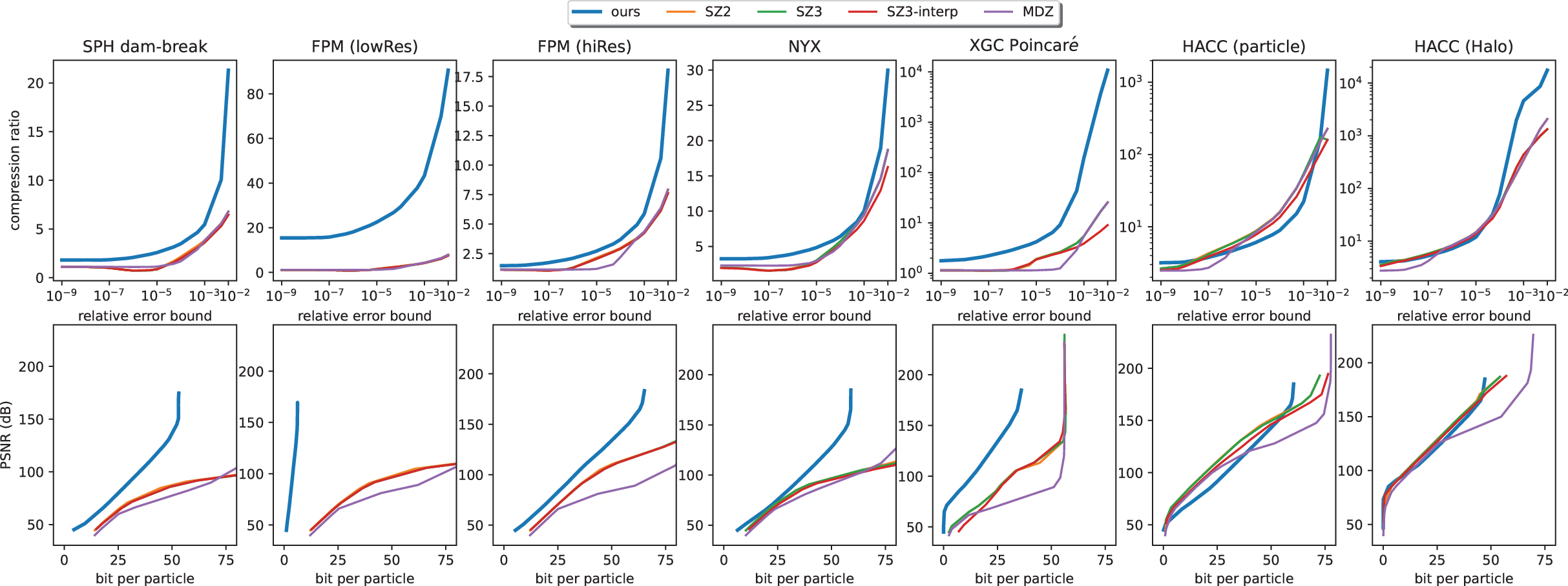}
    \caption{
    Comparison of compression ratios and rate distortions among our method, SZ2, SZ3 (with default predictor), SZ3 with linear interpolation predictor (shown as ``SZ3-interp'' in legend), and MDZ on different datasets.
    }
    \label{fig:performance}
\end{figure*}

\begin{table}[!th]
    \centering
    \scriptsize
    \caption{Benchmark datasets. The three resolution levels of FPM datasets are attained through the manipulation of a parameter known as the smoothing length.}
    \begin{tabular}{c|c|c|c}
    \toprule
        dataset & dim & \# nodes & precision \\ \midrule
        SPH dam-break & 3D & 80,703 & single \\
        Nyx & 3D & 262,144 & double \\
        FPM (lowRes) & 3D &  197,761 & single \\
        FPM (midRes) & 3D & 571,813 & single \\
        FPM (hiRes) & 3D & 1,824,971 & single \\
        XGC \poincare & 2D & 13,077,313 & single \\
        HACC (particle) & 3D & 2,506,573 & double \\
        HACC (Halo downsampled) & 3D & 2,612,364 & double \\
        HACC (Halo) & 3D & 12,219,287 & double \\
    \bottomrule
    \end{tabular}
    \label{tab:datasets}
\end{table}

\begin{figure}[!th]
    \centering
    \includegraphics[width=\linewidth]{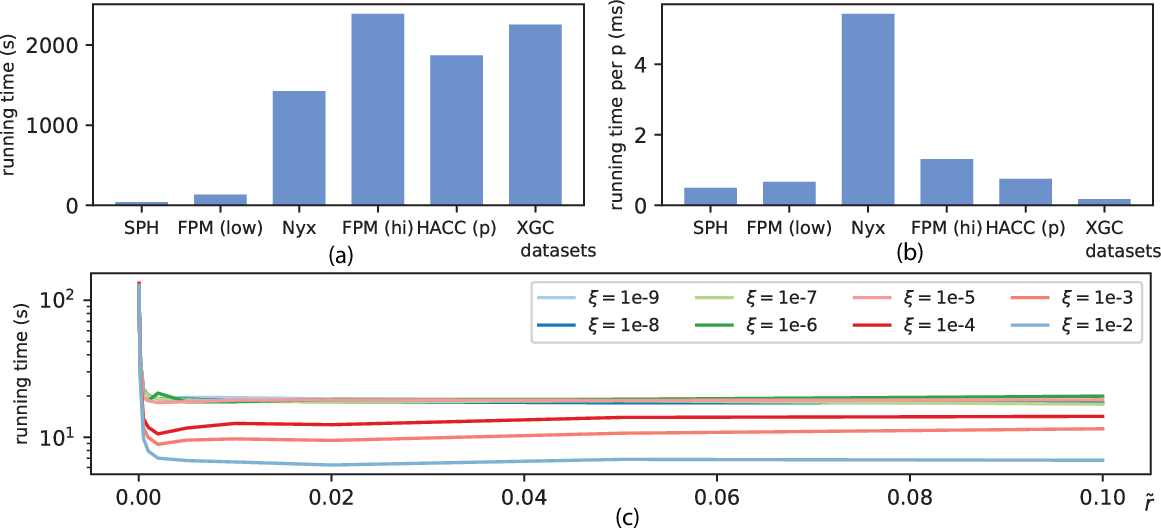}
    \caption{Running time for our compression algorithm. (a) and (b) show the total running time and the running time per particle for six datasets with $\xi=10^{-9}$ and $\tilde{r}=10^{-5}$, respectively, with datasets ordered by the number of particles. (c) illustrates the running time on FPM (lowRes) dataset with varying $\xi$ and $\tilde{r}$.}
    \label{fig:runtime}
\end{figure}

\section{Evaluation}
\label{sec:eval}

We conduct a comparative analysis of our method with several state-of-the-art lossy compressors, including SZ2, SZ3 with default predictor, SZ3 with linear interpolation predictor, and MDZ.
SZ2 uses the Lorenzo predictor~\cite{ibarria2003out}, which is typically superior to linear interpolation for predicting data in a 1D array. Since SZ2 and SZ3 (collectively called SZ) are tailored for regular grid data, the challenge arises when reorganizing particle data, which inherently possess irregular structures, while maintaining a high degree of spatial coherence to adapt SZ to particle data.
Previous studies compressing HACC data with SZ treat particle coordinates along each dimension as a 1D array ordered by their in-memory indices and apply separate compression to different dimensions~\cite{tao2017depth}. We adopt this approach to evaluate SZ. MDZ is designed for molecule dynamics simulation datasets where every data point is an atom, thus suitable for particle data.
All experiments are conducted using a 2021 MacBook Pro equipped with an Apple M1 CPU and 64 GB of main memory.

Our benchmark datasets are specified in~\Cref{tab:datasets}. Smoothed Particle Hydrodynamics (SPH) dam-break data models the release of a large volume of fluid caused by dam rupture via Lagrangian particle method in which the fluid is discretized by a set of particles~\cite{sun2019study}.
Nyx data is generated from cosmological simulations carried out with the Nyx code~\cite{almgren2013nyx}. FPM ensemble datasets from SciVis contest 2016~\cite{scivisContest} model by Finite Pointset Method (FPM) the dynamic mixing phenomenon occurring within a water-filled cylinder, subjected to the infusion of salt solutions from an infinitely supplied salt source at the top. XGC \poincare data is derived by a \poincare plot~\cite{brennan2001existing} that is constructed by plotting the intersection points of plasma particles' trajectories simulated by X-point Gyrokinetic Code (XGC)~\cite{chang2008spontaneous} with a cross section at some toroidal angle in cylindrical coordinate system. HACC datasets refer to large cosmological simulations conducted using the Hardware/Hybrid Accelerated Cosmology Code (HACC)~\cite{habib2016hacc,heitmann2019hacc}. We use HACC datasets from the Mira-Titan Universe Simulations~\cite{hacc_data,heitmann2016mira}, including simulation particles, downsampled Halo particles, and full Halo particles.

\begin{figure*}[!th]
    \centering
    \includegraphics[width=\textwidth]{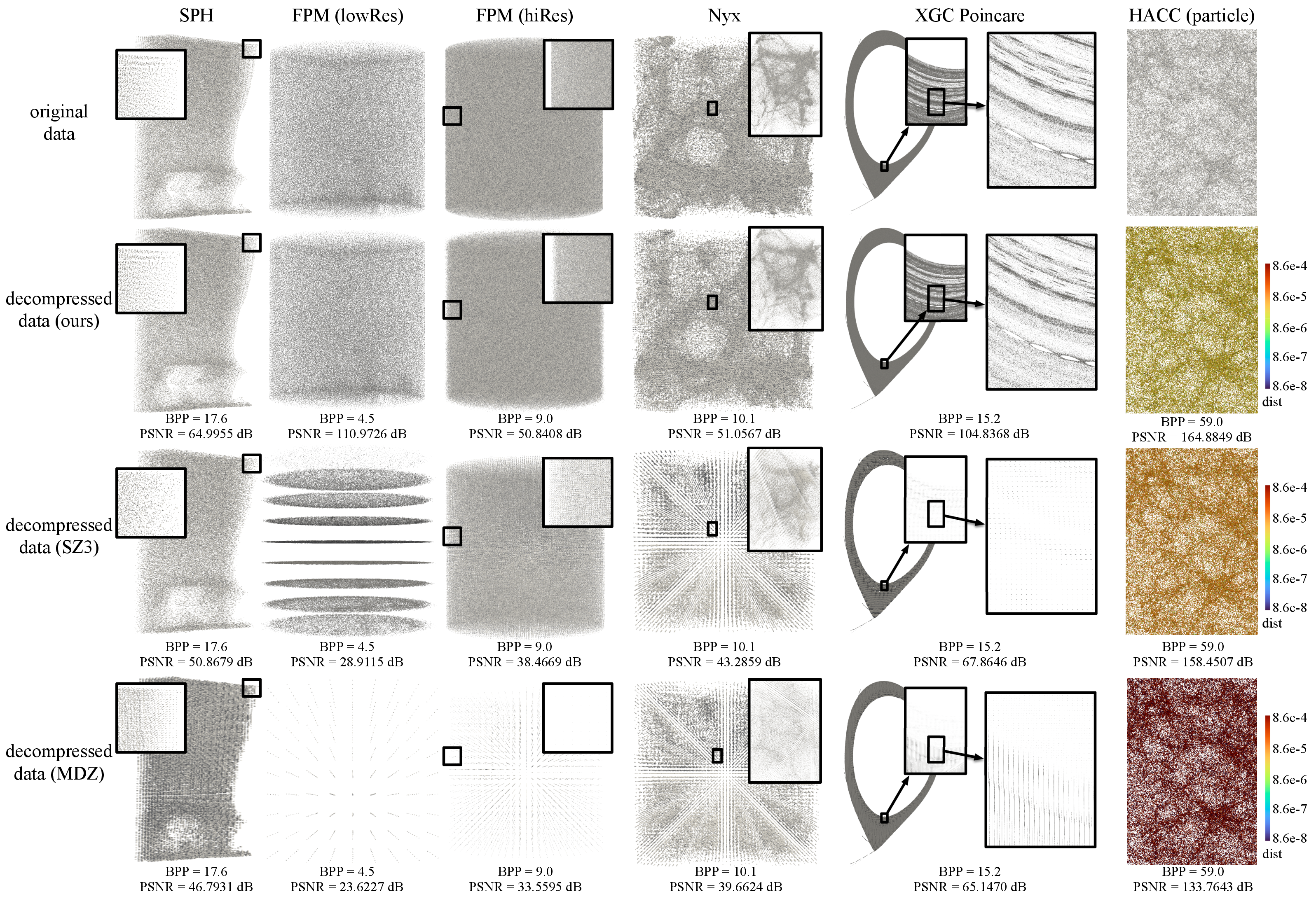}
    \caption{Particle volume rendering results with the same BPP of our method, SZ3, and MDZ on different datasets and their attributes. Specifically, we encode pointwise $l_2$ distance by color in visualizations of decompressed HACC (particle) data because the values of $\xi$ ($10^{-8}$ for our method, $2.4\times10^{-8}$ for SZ3, and $3\times10^{-7}$ for MDZ) are too small to be visible with 2,506,573 particles.}
    \label{fig:rendering}
\end{figure*}

We quantitatively evaluate the performance of our method across four dimensions: (1) compression ratio vs. $\xi$, where $\xi$ is the relative error bound normalized by the max range of original coordinates in all dimensions (\Cref{fig:performance} first row);
(2) normalized root mean squared error (NRMSE) vs. $\xi$ (Supplemental Materials (SMs)); (3) rate distortion, i.e., peak signal-to-noise ratio (PSNR) vs. bits per particle (BPP) (\Cref{fig:performance} last row); (4) running times of compression and decompression process vs. $\xi$ (\Cref{fig:runtime}). Specifically, we present the results for the value of $\tilde{r}$ that yields the best compression ratio. Then, we conduct a qualitative comparison of the rendering results achieved by our method and two baseline approaches under the same PSNR (\Cref{fig:rendering}). Furthermore, we analyze the best setting of $\tilde{r}$, and perform an ablation study to dissect the impact of reordering particles and dynamically selecting the number of bits to encode a quantization code.

\begin{figure}[!t]
    \centering
    \includegraphics[width=\linewidth]{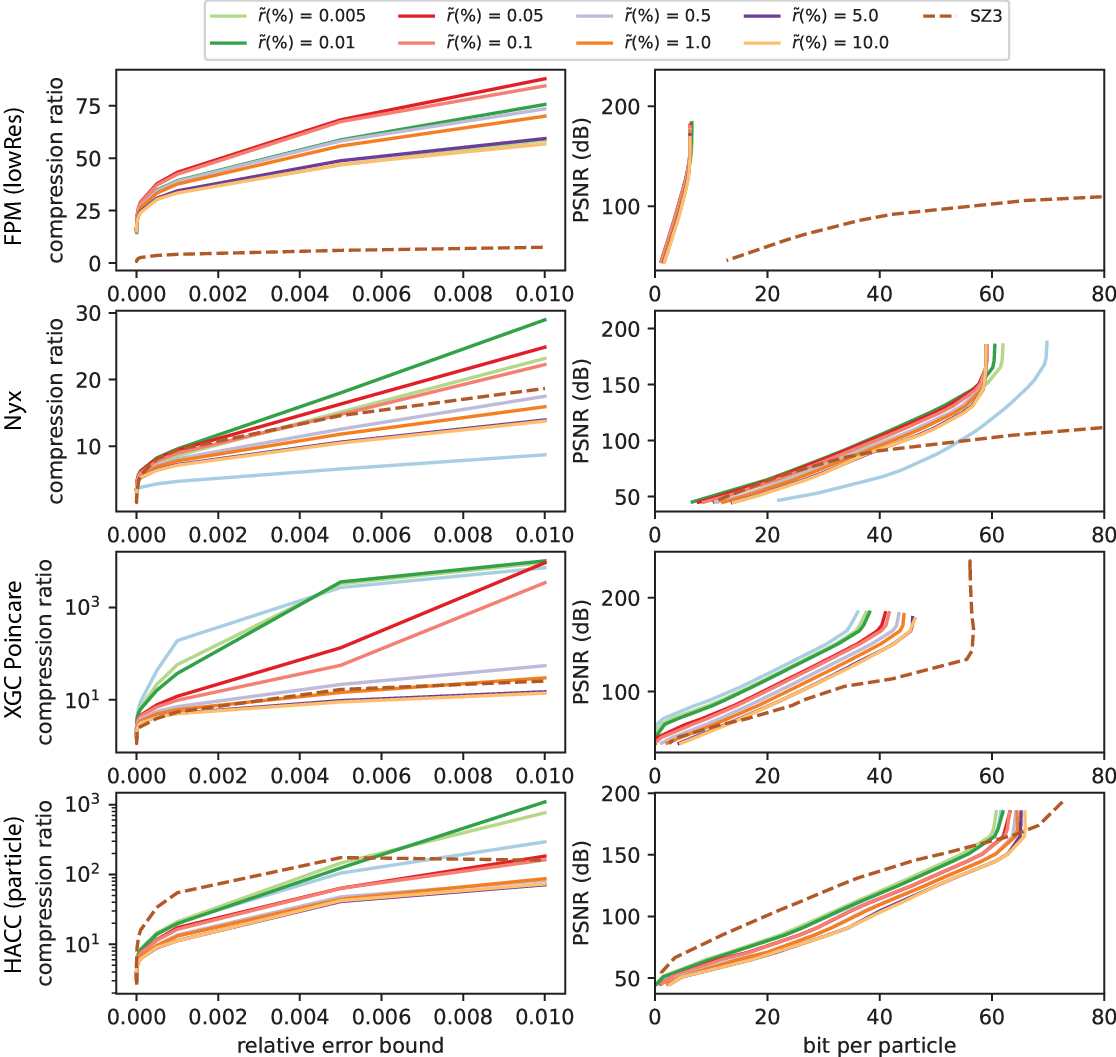}
    \caption{Compression ratios and rate distortion under different $\tilde{r}$'s for four datasets, FPM (lowRes), Nyx, XGC \poincare, and HACC (particle).}
    \label{fig:parameter_analysis}
\end{figure}

\subsection{Baseline Comparison}
\label{sec:baseline}
We compare the compression ratios, NRMSEs, and rate distortions of our method to those of SZ and MDZ. Additionally, we examine the factors influencing running time and evaluate the visualization quality of decompressed data produced by different compression algorithms. The results of NRMSEs for all datasets and compression ratios and rate distortions on two omitted datasets are shown in SMs.

\textbf{Compression ratio vs. $\xi$}. We examine a range of relative error bounds that extends from $10^{-9}$ to $10^{-2}$. We achieve a superior compression ratio compared to other compressors for most of the datasets tested, as illustrated in~\Cref{fig:performance}. In the case of HACC datasets, our method exhibits higher compression ratios than other compressors when the error bounds are very low. For example, our method exhibits at least a 1.2$\times$ compression ratio improvement over other compression methods for $\xi=10^{-9}$
on the HACC particle data. This suggests that our method is particularly beneficial when prioritizing super accuracy.

MDZ is not suitable for the datasets we tested. MDZ targets molecular dynamics datasets with multiple snapshots, characterized patterns (e.g., zigzag or stairwise patterns) in the multi-level distribution in the spatial domain, and high smoothness of successive snapshots along the temporal dimension. However, we tested non-time-varying datasets or one instance from time-varying datasets without similar spatial or temporal distribution properties.

\textbf{NRMSE vs. $\xi$}. NRMSE and PSNR are calculated by the following equations for 3D datasets, respectively:
\begin{equation*}
    NRMSE(p,\hat{p})=\sqrt{\frac{1}{3N}\sum_{i=0}^{N-1}\sum_{d\in\{x,y,z\}}\left(\frac{p_{i,d}-\hat{p}_{i,d}}{\Delta_{max}}\right)^2}\mbox{, and}
\end{equation*}
\begin{equation*}
    PSNR(p,\hat{p})=-20log_{10}NRMSE(p,\hat{p})\mbox{,}
\end{equation*}
where $N$ is the number of particles; $p_{i,d}$ and $\hat{p}_{i,d}$ represent the original and decompressed coordinates of the $i$-th particle in dimension $d$, respectively. Our method reveals very similar NRMSE compared with SZ-family compressors, while MDZ exhibits NRMSE very close to the upper bound with relative error bounded, as shown in SMs.

\textbf{Rate distortion (PSNR vs. BPP)}. We achieve higher PSNR under the same BPP for most of tested datasets, except HACC datasets. For HACC particle and HACC Halo datasets, our method exhibits higher PSNR values in high BPP range (around or greater than 50), which again indicates its utility when super accuracy is required.

\textbf{Running time}. The time complexity of our compression method is $O(|B|log|B|)$, which is highly correlated to $\tilde{r}$ because $|B|\cdot \tilde{r}\approx 1$ and seems to be independent of $\xi$. However, $\xi$ affects the lengths of bit boxes (as shown by Equations~\ref{eq:num_bits} and~\ref{eq:box_length}), and thus affects the number of bit box intersections, the number of common particles in the intersection of bit boxes, and the number of sequences, which slightly impacts the running time. We plot the running time vs. $\tilde{r}$ for some of $\xi$ on FPM (lowRes) data in~\Cref{fig:runtime}(c), which shows that the running time is dominated by $\tilde{r}$ while slightly differs under different $\xi$'s. Generally, the running time increases with the number of particles in the dataset (\Cref{fig:runtime}(a)). However, the time to process a particle varies by dataset (\Cref{fig:runtime}(b)), which is dominated by the number of bit box intersections. We omit comparing running time with other compressors because our Python implementation may not be as optimized as their C/C++ implementations, which is left as a topic for future study.

\textbf{Visual qualities}. We compare rendering results of decompressed data produced by our method, SZ3 (with default predictor), and MDZ under the same BPP (\Cref{fig:rendering}). Rendering results for SZ2 and SZ3 with linear interpolation predictor are omitted here as they are similar to those for SZ3. Our method produces visualizations with better PSNR and visual quality, particularly on the FPM (lowRes), Nyx, and XGC \poincare datasets. For the FPM (lowRes) dataset, the decompressed data by SZ3 and MDZ fail to capture the cylindrical structure of the original data; one exhibits several layers, while the other shows 3D lines. In the case of the Nyx dataset, SZ3 and MDZ introduce axis-aligned patterns that do not exist in the original data. Regarding the XGC \poincare dataset, our method preserves the island structure around a saddle point on the separatrix of the magnetic flux field (refer to the second-level detailed figure), which is crucial for analysis but lost in the reconstructed data by SZ3 and MDZ. For the HACC (particle) dataset, which demands high accuracy in analysis, we encode the $l_2$ distance pointwise by color to illustrate nuances that are hardly discernible based solely on particle distribution with a limited number of particles.

\subsection{Parameter Analysis}
\label{sec:parameter_analysis}

We analyze the effect of the hyperparameter, $\tilde{r}$, which indicates the ratio of maximum particle counts in a subregion to the total particle counts, on compression ratio and rate distortion. The results under different $\tilde{r}$'s, along with SZ3 with the default predictor setting as baseline, on several datasets are presented in~\Cref{fig:parameter_analysis}. Other baseline compressors used in~\Cref{sec:baseline} exhibit similar or even worse rate distortion and are thus omitted in the parameter analysis. The results for other datasets are provided in SMs. There are three observations concerning the hyperparameter settings, as listed below.

\textbf{O1}. Whether our method outperforms existing compressors depends on the proper setting of $\tilde{r}$. Although the value of $\tilde{r}$ does not affect rate distortion significantly on some datasets (e.g., FPM (lowRes)), for most tested datasets, including Nyx, XGC \poincare, and HACC (particle), SZ3 exhibits better compression ratios and rate distortion than several values of $\tilde{r}$ do, especially for PSNR in the low BPP range. Additionally, neither large nor small $\tilde{r}$ values yield the best results. Taking Nyx dataset as an example, both the largest and smallest parameter settings (i.e., $\tilde{r}=10\%$ and $\tilde{r}=0.001\%$, respectively) show lower compression ratios than SZ3 does. Large $\tilde{r}$ values exhibit poor performance because they fail to fit the particle density optimally. Ideally, small bit boxes should be generated for high-density areas, whereas large bix boxes tend to mix high- and low-density areas. Conversely, small $\tilde{r}$ values incur too many bit boxes, increasing the number of bit box centers that need losslessly storing and resulting in large storage overhead.

\textbf{O2}. The optimal value of $\tilde{r}$ is a function of $\xi$.
This observation is best illustrated by the results on XGC \poincare dataset. A small value of $\tilde{r}$ (0.001\%)
exhibits the highest compression ratio in the low error bound range, while a higher value of $\tilde{r}$ (0.01\%) outperforms the former as $\xi$ increases. Referring to Equations~\ref{eq:num_bits} and~\ref{eq:box_length}, $\xi$ affects the length of bit boxes due to the ceiling function, which ensures the size of quantization codes is an integer. The larger $m_d$ is raised by the ceiling function, the more significant the effect of $\xi$ on the length of a bit box becomes. Thus, different settings of $\xi$ change the performance ranking of $\tilde{r}$ values, but the optimal $\tilde{r}$ does not vary significantly as $\xi$ increases.

\textbf{O3}. The optimal value of $\tilde{r}$ is more closely related to the spatial distribution of particles rather than the data size alone. Despite the increasing orders of particle counts in Nyx, HACC (particle), and XGC \poincare datasets, the optimal value of $\tilde{r}$ remains at 0.01\% in the high error bound range. This suggests that there is no universally optimal number of particles that a bit box should accommodate. As shown in~\Cref{fig:rendering}, the particle distributions of the three datasets differ a lot. The XGC \poincare dataset, with the highest number of particles, exhibits an avocado-like shape and highly nonuniform spatial distribution. Thus, the high-density areas could be separated by relatively large bit boxes. On the contrary, the Nyx dataset displays a more uniform distribution than the XGC \poincare dataset, requiring more subregions and smaller bit boxes to distinguish between high- and low-density areas.

Based on the datasets we tested, we recommend starting with $\tilde{r}$ values by including $4\times 10^3$ to $9\times 10^3$ particles in a subregion. Determining a theoretically optimal $r$ value would require assumptions on the spatial distribution of particles, which are not made in this paper. Prior knowledge of the particle distribution, along with more specific particle partitioning strategies, would be beneficial for a specific dataset. This avenue remains for future exploration.

\begin{figure}[!th]
    \centering
    \includegraphics[width=\linewidth]{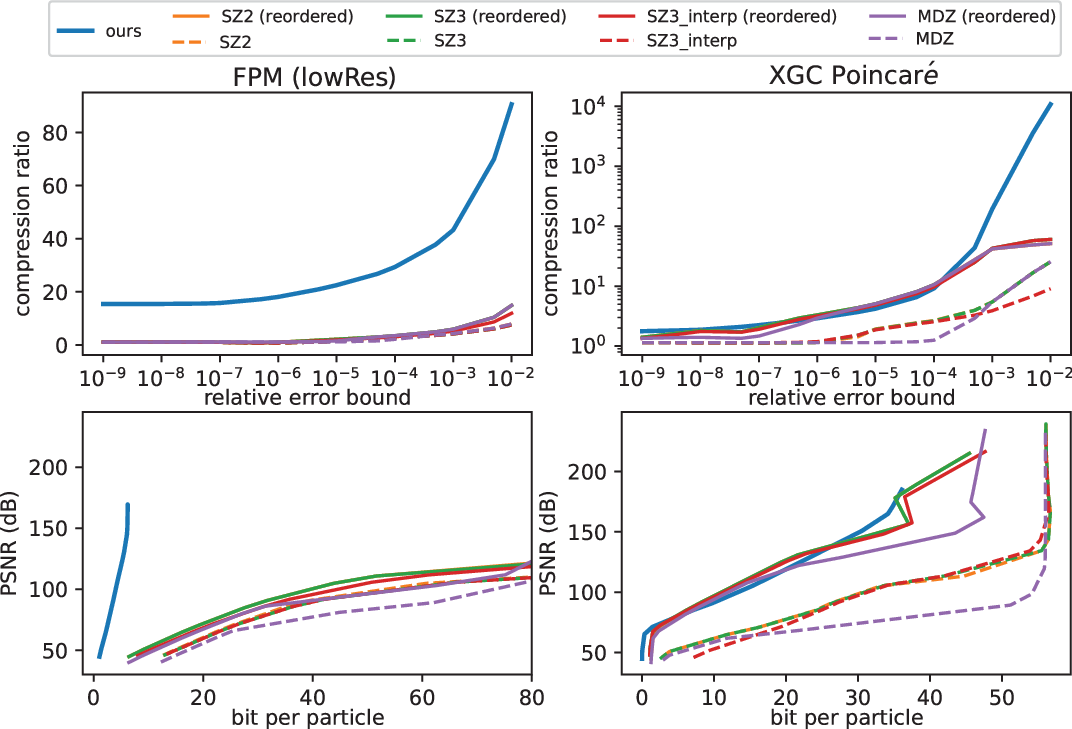}
    \caption{Ablation study on FPM (lowRes) and XGC \poincare datasets. The results for compressors with input data organized by in-memory ordering are shown in dashed lines while those with input data reordered by our data layout in~\Cref{fig:sequences} are presented in solid lines.
    }
    \label{fig:ablation}
\end{figure}

\subsection{Ablation Studies}
We conduct an additional experiment to investigate the impact of the two mechanisms in our method, which couples reordering particles by bit box containment to enhance spatial coherence between particles with successive indices, and dynamically adjusting the lengths of bit boxes to minimize the bits used for quantization codes. We test the following hypotheses to break down the effects of the two mechanisms:\\
\textbf{H1}. Our reordering technique enhances the performance of regular-grid-based compressors, such as SZ.\\
\textbf{H2}. The two coupled mechanisms in our method, reordering particle indices by spatial grouping and adaptively adjusting the lengths of quantization codes, both contribute to the improvement of performance.\\
To test the two hypotheses, we compare (1) our method, (2) SZ and MDZ, and (3) SZ and MDZ taking the particles reordered as data layout shown in~\cref{fig:sequences}. The results for two datasets are presented in~\Cref{fig:ablation}. Additional results for other datasets can be found in SMs.

\textbf{Results}. The results obtained from both datasets provide strong support for \textbf{H1}. SZ and MDZ with reordered particles exhibit significant enhancements in both the compression ratio (the first row in~\Cref{fig:ablation}) and rate distortion (the last row in~\Cref{fig:ablation}), compared to them with in-memory ordered particles. However, \textbf{H2} is only partially manifested. Comparing our method that couples two mechanisms and MDZ with reordered particles, we can see that the coupled mechanism may not consistently yield the best performance. Our method outperforms SZ and MDZ with reordered particles on the FPM (lowRes) dataset. In contrast, on the XGC \poincare dataset, the prediction by the center with an adaptive number of bits does not surpass the predictor of MDZ.
\section{Limitations and Future Works}
\label{sec:discussion}

\textbf{Limitations}. While our compression algorithm is not limited to 2D or 3D particle data, as demonstrated in~\Cref{sec:eval}, several open topics warrant further investigation. First, the strategy to predetermine the optimal parameter $\tilde{r}$ remains unclear. Determining the optimal $\tilde{r}$ poses challenges due to the unknown distribution of particles. For instance, a uniform distribution results in fewer intersections of bit boxes, while clustered particles lead to more intersections, complicating the estimation of storage costs for the data layout depicted in~\Cref{fig:sequences}. Also, predicting the size reduction by Huffman encoding and ZSTD compression is challenging. While the entropy of input data provides insight into the converged reduction size of Huffman encoding, approximating this convergence with a limited number of sequences is impractical. This issue also makes it hard to adaptively adjust $\tilde{r}$ for a (sub)region.

The second limitation concerns the predictor for lossily storing particles. Currently, we employ a straightforward approach where the center is used to predict all the particles within a bit box. However, other prediction methods, such as those using predicted values as well as losslessly saved values to predict new values, have the potential to reduce prediction error and subsequently encode more particles with fewer bits. Exploring and implementing such wiser prediction strategies could enhance the efficiency of our compression algorithm.

\textbf{Future works}. A promising direction for future research is exploring alternative clustering methods beyond the k-d tree. While our k-d splitting technique based on the median depicts particle density by adjusting the lengths of bit boxes, it does not ensure that the centers of bit boxes align with the centroids of clusters. For example, in a dataset comprising two clusters with 100 particles in total, where one cluster contains 70 particles and the other 30, k-d splitting would divide them into 50 particles on each side, disrupting one cluster and complicating prediction tasks. Investigating clustering methods that better preserve the spatial distribution of particles and align bit box centers with cluster centroids could improve compression performance.

Our method also holds the potential for adaptation to other applications. First, we can trim our method to support progressive compression and decompression, where the loss of particles is deemed acceptable. In progressive compression, we adopt a sampling strategy that involves selecting more particles within smaller bit boxes and fewer particles within larger ones. This approach can also be extended to the progressive decompression process. Second, for datasets including multiple instances wherein particles exhibit localized variations, such as members within an ensemble or timesteps in time-varying data, we can identify the bit boxes where these variations occur across two successive instances. Then, it is easy to dynamically update the bit boxes and their intersections with small computational overhead. Third, our method can support compressing continuous attributes of particles, such as velocity. Straightforward methods include reusing the bit boxes constructed for position data to reduce storage for losslessly saving centers and constructing another set of bit boxes for attributes to minimize the length of quantization codes. We leave the comparison and proposal of generalization approaches open for future investigation.
\section{Conclusion}
\label{sec:conclusion}

We propose an error-bounded lossy compression technique tailored specifically for particle position datasets originating from a wide array of scientific applications, including cosmology, fluid dynamics, and fusion energy sciences. Inspired by quantization-encoding schemes in SZ, our method groups the particles and dynamically determines the optimal number of bits to encode each group of the particles, ensuring error control while minimizing the bit count used for compression. Through both quantitative and qualitative comparison with several state-of-the-art compressor, SZ2, SZ3 with different predictor settings, and MDZ, we demonstrate the effectiveness and efficiency of our approach in achieving high compression ratios while maintaining error control and data fidelity across diverse datasets. This work lends itself to several potential extensions. First, further exploration could focus on predetermining parameters to optimize compression ratios or rate distortions. Second, there is potential for refinement of the algorithm to support progressive compression and decompression, enabling more efficient handling of data with varying levels of detail. Lastly, there is scope for the development of dynamic updating mechanisms tailored to ensemble or time-varying datasets, enhancing the adaptability of the compression method to evolving data structures.

\acknowledgments{
This research is supported by DOE DE-SC0022753, NSF OAC-2311878, NSF OAC-2313123, and NSF IIS-1955764. This research is also supported by the Exascale Computing Project (ECP), project number 17-SC-20-SC, a collaborative effort of the U.S. Department of Energy Office of Science and the National Nuclear Security Administration. It is also supported by the U.S. Department of Energy, Office of Advanced Scientific Computing Research, Scientific Discovery through Advanced Computing (SciDAC) program of the U.S. Department of Energy under Contract No. DE-AC02-06CH11357.
}

\bibliographystyle{abbrv-doi-hyperref}

\bibliography{references}

\end{document}